\documentclass[conference,letterpaper]{IEEEtran}
\usepackage[utf8]{inputenc}
\usepackage[T1]{fontenc}
\usepackage{url}
\usepackage{ifthen}
\usepackage{cite}
\usepackage{xcolor}
\usepackage[cmex10]{amsmath} % Use the [cmex10] option to ensure complicance
\usepackage{cite}
\usepackage{array}
\usepackage{fixltx2e}
\usepackage{stfloats}
\usepackage{url}% Basically, \url{my_url_here}.
\usepackage{amsmath,amssymb}
\usepackage{epsfig,fancyhdr}% Use fancyhdr.sty
\usepackage{CJKutf8}
\usepackage{times}
\usepackage{graphicx}
\usepackage{soul}

\newcommand{\bm}[1]{\mbox{\boldmath $#1$}}
\newcommand{\bms}[1]{\mbox{\footnotesize\boldmath$#1$}}
\newcommand{\bmss}[1]{\mbox{\tiny\boldmath$#1$}}

\newtheorem{thm}{Theorem}
\newtheorem{rmk}{Remark}

\newtheorem{defin}{Definition}

\newtheorem{eg}{Example}

\begin{document}
\title{Sparse Complementary Pairs with Additional Aperiodic ZCZ Property}

\author{
\IEEEauthorblockN{Cheng-Yu~Pai$^{\scriptscriptstyle \dag\ddag}$, Zilong~Liu$^{\scriptscriptstyle \ddag}$, Chunlei~Li$^{\scriptscriptstyle \S}$, and Chao-Yu~Chen$^{\scriptscriptstyle \P}$}
\IEEEauthorblockA{$^{\scriptscriptstyle \dag}$Department of Engineering Science, National Cheng Kung University, Taiwan\\
$^{\scriptscriptstyle \ddag}$School of Computer Science and Electronic Engineering, University of Essex, UK\\
$^{\scriptscriptstyle \S}$Department of Informatics, University of Bergen, Norway\\
$^{\scriptscriptstyle \P}$Department of Electrical Engineering and Institute of Computer and Communication Engineering,\\ National Cheng Kung University, Taiwan}
Email: \{$^{\scriptscriptstyle \dag}$n98081505, $^{\scriptscriptstyle \P}$super\}@mail.ncku.edu.tw  $^{\scriptscriptstyle \ddag}$zilong.liu@essex.ac.uk $^{\scriptscriptstyle \S}$Chunlei.Li@uib.no}

%\author{
%\IEEEauthorblockN{Author 1, Author 2, Author 3, and Author 4}
%}

\maketitle
%\footnotetext[1]{This work was supported in part by the Research Council of Norway under Grant 311646/070, by the Royal Society International Exchange 2022 Cost Share (NSTC) under Grant IEC$\backslash$R3$\backslash$223079, and by the National Science and Technology Council, Taiwan, R.O.C., under Grants NSTC 109-2628-E-006-008-MY3 and NSTC 112-2927-I-006-503}

\begin{abstract}
  %To enable low-complexity, low-latency, and low-storage signal processing in modern communication systems,
   This paper presents a novel class of complex-valued sparse complementary pairs (SCPs), each consisting of a number of zero values and with additional zero-correlation zone (ZCZ) property for the aperiodic autocorrelations and crosscorrelations of the two constituent sequences. Direct constructions of SCPs and their mutually-orthogonal mates based on restricted generalized Boolean functions are proposed. It is shown that such SCPs exist with arbitrary lengths and controllable sparsity levels, making them a disruptive sequence candidate for modern low-complexity, low-latency, and low-storage signal processing applications.   %We also study sparse complementary mates (SCMs) which have zero aperiodic crosscorrelation sums.

\end{abstract}

\section{Introduction}
The concept of Golay complementary pair (GCP) \cite{Golay} refers to a pair of sequences with zero aperiodic autocorrelation sum (AACS) at every non-zero time-shift. Driven by the diverse applications of GCPs in coding and telecommunication \cite{Golay_RM,Chen_CDMA,Liu_2015,Liu_2014,Liu_2019,CS_sync,Wang2007,Pezeshki_08}, there are also Golay complementary sets (GCSs) \cite{Tseng72,Paterson_00,Super_16,schmidt}, each comprised of two or more constituent sequences, and Z-complementary pairs (ZCPs) with zero-correlation zone (ZCZ) sum properties\cite{ZCP-1st,even-ZCP,odd-ZCP,Super_17,Adhikary_19, Pai_202}.
%Besides, some research has further studied GCPs with zero periodic correlation zone properties \cite{Gong13,Super_182,Gu_21} which can be used as training sequences to improve the performance of channel estimation.

With numerous research works on complementary sequence pairs/sets, the existing state-of-the-art mostly does not consider the auto- and cross-correlations among the constituent sequences. This is perhaps because of the common assumption that these constituent sequences are sent over orthogonal frequency/time channels. However, in a practical communication scenario where the transmission of one sequence interferes with the other, good inter-sequence aperiodic correlation properties are highly desirable. Among few exceptions, cross Z-complementary pairs (C-ZCPs) were first proposed in 2020 for optimal channel training in broadband spatial modulation systems \cite{Liu_2020}, where each pair is characterized by the ZCZ properties for both their AACSs and aperiodic crosscorrelation sums. For Doppler-resilient radar waveforms, an efficient majorization minimization algorithm was developed in \cite{Wang_21} for sequence pairs called quasi-orthogonal ZCPs (QO-ZCPs). In a QO-ZCP, both the AACSs and aperiodic crosscorrelation values (rather than the aperiodic crosscorrelation sums) of the two constituent sequences have near-zero values within the zone.

The main objective of this work is to introduce a novel class of complex-valued sparse complementary pairs (SCPs), where the constituent sequences of each SCP contain a number of zeros and exhibit zero aperiodic auto- and cross-correlation zone property\footnote{Although there are ternary complementary pairs (TCPs) \cite{Gavish_94,Craigen_01} over the alphabet of $\{+1,0,-1\}$, the aperiodic correlation properties of the constituent sequences have not been investigated.}. These sparse pairs are useful in signal processing and  communication/radar system design. For example, with the recent advances in sparse signal processing, the sequence sparsity can be smartly exploited for low-complexity, low-latency, and low-storage hardware implementation \cite{Tinney_85,Cao_09,Cao_11,Zhang_15}. For the aforementioned channel training of spatial modulation and Doppler-resilient radar sensing, such sparse pairs may be used as an alternative for C-ZCPs and QO-ZCPs, respectively.

Besides, we study the sparse complementary mate of an SCP, where the two pairs are mutually orthogonal in terms of the zero aperiodic cross-correlation sums for all the time shifts. It is shown that through a careful design, the four constituent sequences (see ${\bm C}_0,{\bm C}_1,{\bm S}_0,{\bm S}_1$ in Fig. \ref{fig:SCM}) can also enjoy zero aperiodic cross-correlation zone property. By using restricted generalized Boolean functions (RGBFs) \cite{Paterson_00,schmidt}, the core idea of our proposed constructions is to carefully restrict certain variables of RGBFs, thus leading to flexible sequence lengths, sparsity, and ZCZ widths.  %It is worth mentioning that sequences from RGBFs have algebraic structure and hence can lead to faster code generation.

The rest of this paper is outlined as follows. Section \ref{sec:background} introduces some notations and preliminaries including the definitions of SCPs and their mates. The concept of RGBFs is introduced in Section \ref{sec:RGBF}. Then we present direct constructions of SCPs and SCPs based on RGBFs in Section \ref{sec:SCP}. Finally, we conclude our paper in Section \ref{sec:conclusion}.
\section{Background and Definitions}\label{sec:background}
The following notations will be used throughout this paper:
\begin{itemize}
%\item $\bm 1$ is an all-one vector.
\item $\mathbb{Z}_q=\{0,1,\ldots,q-1\}$ denotes the set of integers modulo a positive integer $q$.
\item $\xi=e^{2\pi \sqrt{-1}/q}$ denotes a $q$-th primitive root of unity.
\item $(\cdot)^{*}$ denotes the complex conjugation.
%\item ${\bm 1}$ represents for an all-one vector.
\end{itemize}

Let ${\bm C}_0$ and ${\bm C}_1$ be two complex-valued sequences  of length $L$ given by
% \begin{equation}
% \begin{aligned}
${\bm C}_k=(C_{k,0},C_{k,1},\ldots,C_{k,L-1}),$
% \end{aligned}
% \end{equation}
where $|C_{k,i}|\in \{0,1\}$ and ${C}_{k,0},{C}_{k,L-1}\neq 0$.
The aperiodic cross-correlation of sequences ${\bm C}_0$ and ${\bm C}_1$ at the time-shift $u$ is defined as
\begin{equation}
\rho({\bm C}_0,{\bm C}_1;u)=
\begin{cases}
\sum_{i=0}^{L-1-u}C_{0,i+u}C_{1,i}^{*},&  0\leq u<L;\\
\sum_{i=0}^{L-1+u}C_{0,i}C_{1,i-u}^{*},&  -L< u<0.
\end{cases}
\end{equation}
Note that $\rho({\bm C}_0,{\bm C}_1;u)=\rho^{*}({\bm C}_1,{\bm C}_0;-u)$. If ${\bm C}_0={\bm C}_1$, $\rho({\bm C}_0,{\bm C}_0;u)$ is called the aperiodic autocorrelation of ${\bm C}_0$, denoted as $\rho({\bm C}_0;u)$.
In what follows, we formally define {\it sparse complementary pairs}.
\begin{defin}
Let $({\bm C}_0,{\bm C}_1)$ be a pair of complex-valued sequences with length $L$ and $N$ zero elements in each sequence. Suppose that the following two conditions hold:
\begin{equation}\label{eq:SCP}
\begin{array}{l}
\hspace{-10pt}{\text C1}:\rho({\bm C}_{k},{\bm C}_{k'};u)
=\begin{cases}
L-N, &  u=0,~k=k'\\
0, &  0<|u|< Z,k=k'\\
0, &  |u|< Z,~k\neq k'
\end{cases}
\\ [0.2cm]
\hspace{-10pt}{\text C2}:\rho({\bm C}_0;u)+\rho({\bm C}_1;u)=\begin{cases}
0, &  u\neq 0\\
2(L-N), &  u=0.
\end{cases}
\end{array}
\end{equation}
Then $({\bm C}_0, {\bm  C}_1)$ is said to be an $(L,Z,\mathcal{S})$-SCP, where $Z$ denotes the width of ZCZ and $\mathcal{S}=N/L$ represents the sparsity level.
\end{defin}

\begin{figure}[t!]
	\centering
	\includegraphics[width = 3.5in]{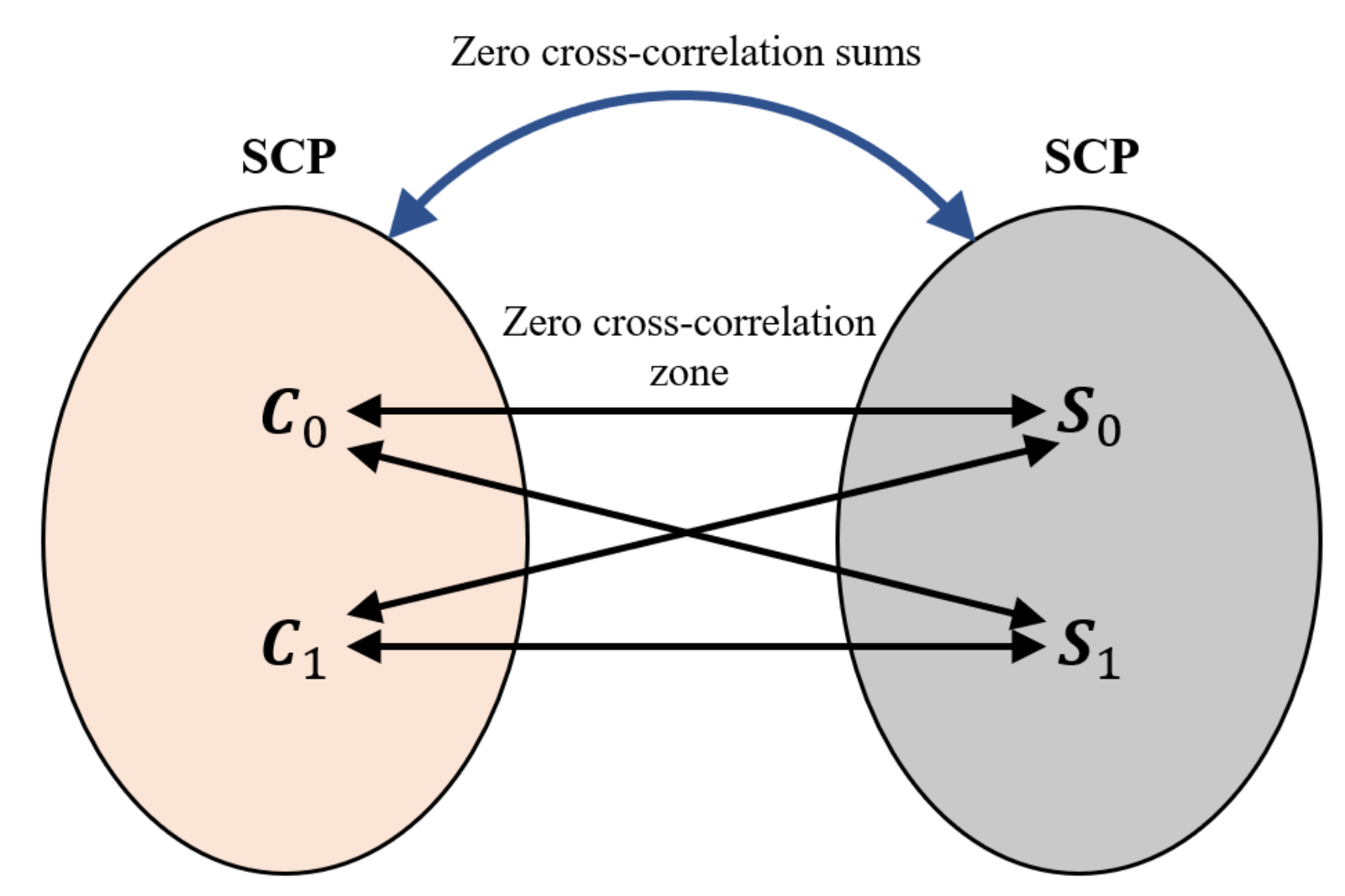}
	\caption{Aperiodic cross-correlation properties of an SCP and its mutually orthogonal mate.}
    \label{fig:SCM}	
\end{figure}

Note that for the well-known GCP ($\mathcal{S}=0$), only C2 in the above definition is satisfied.
\begin{defin}
Two distinct $(L,Z,\mathcal{S})$-SCPs $({\bm C}_0,{\bm C}_1)$ and $({\bm S}_0,{\bm S}_1)$ are said to be the mate  of each other if
\begin{equation}\label{eq:scm}
\begin{aligned}
&\text{C1:}~\rho({\bm C}_{0},{\bm S}_{0};u)+\rho({\bm C}_{1},{\bm S}_{1};u)=0,~  |u|<L;\\[0.2cm]
&\text{C2:}~\rho({\bm C}_{k},{\bm S}_{k'};u)=0,~  |u|< Z,~\text{and}~ k,k'=0,1.
\end{aligned}
\end{equation}
\end{defin}
The cross-correlation properties of these two SCPs are depicted in Fig. \ref{fig:SCM}.

%\begin{defin}
%For an $(L,Z_A,M/L)$-SCP, the ZCZ ratio is defined as
%\begin{equation}
%ZCZ_{\text{ratio}}=\frac{Z_A}{L}.
%\end{equation}
%The maximum value of the ZCZ ratio is $ZCZ_{\text{ratio}}=\frac{L-1}{L}$.
%\end{defin}
\section{Restricted Generalized Boolean Functions}\label{sec:RGBF}
In this section, we will introduce the concept of restricted generalized Boolean functions (RGBFs) \cite{Paterson_00,schmidt}. First, we introduce the concept of GBF. A GBF $f$: $\mathbb{Z}^{m}_2$ $\rightarrow$ $\mathbb{Z}_q$ is composed of $m$ variables $x_1,x_2,\ldots,x_m$ where $x_l \in \{0,1\}$ for $l=1,2,\ldots,m$. The sequence ${\bm f}$ corresponding to the GBF $f$ is denoted by $\bm f=(f_0,f_1,\ldots,f_{2^{m}-1})$ where $f_i=f(i_1,i_2,\ldots,i_m)$ and $i=\sum_{l=1}^{m}i_{l}2^{l-1}$. The complex-valued sequence associated with ${\bm f}$ is given by
$\xi^{\bms f}=(\xi^{f_0},\xi^{f_1},\ldots,\xi^{f_{2^{m}-1}})$.
\begin{eg}\label{eg:gbf}
Taking $q=4$, $m=3$, the associated $q$-ary sequence ${\bm f}$ to the GBF $f=2x_2x_3+x_1$ is given by ${\bm f}=(01010123)$. The complex-valued sequence is ${\xi^{\bms f}}=(\xi^{0},\xi^{1},\xi^{0},\xi^{1},\xi^{0},\xi^{1},\xi^{2},\xi^{3})$.
%${\xi^{\bms f}}=(\xi^{0}\xi^{1}\xi^{0}\xi^{1}\xi^{0}\xi^{1}\xi^{2}\xi^{3})$.
\end{eg}

Consider a set of $t$ indices $V=\{v_1,v_2,\ldots,v_t\}\subset \{1,2,\ldots,m\}$ and a set of $m-t$ indices $V'=\{v'_1,v'_2,\ldots,v'_{m-t}\}=\{1,2,\ldots,m\}\setminus V$, where $0\leq t<m$. Let ${\bm X}=(x_{v_1},x_{v_2},\ldots,x_{v_t})$ and ${\bm d}=(d_1,d_2,\ldots,d_t)$ where $d_{l}\in \{0,1\}$. A RGBF $f|_{\bmss X={\bms d}}$ is defined by restricting variables ${\bm X}$ in the $f$ to the certain known ${\bm d}$. For simplicity, let the associated sequence ${\bm f}|_{{\bmss X}={\bms d}}$ be the complex-valued sequence with component equal to $\xi^{f_i}$ if $i_{v_{\alpha}}=d_{\alpha}$ for $\alpha=1,2,\ldots,t$, and equal to zero otherwise. Let ${\bm F}=(F_0,F_1,\ldots,F_{2^{m}-1})={\bm f}|_{{\bmss X}={\bms d}}$. Let $k_0$ and $k_1$ be the smallest and the largest integers, satisfying $F_{k_0},F_{k_1}\neq 0$, respectively. It is clear that
\begin{equation}\label{eq:front_end}
k_0=\sum_{\alpha=1}^{t}d_{\alpha}2^{v_{\alpha}-1}~\text{and}~k_1=k_0+\sum_{\alpha=1}^{m-t}2^{v'_{\alpha}-1}.
\end{equation}
Let $L=k_1-k_0+1$ and  define ${\bm f}^{(L)}|_{{\bmss X}={\bms d}}$ as the truncated sequence by removing the first $k_0$ elements and the last $2^{m}-1-k_1$ elements. Therefore, the constructed sparse sequence is of length $L$. In this paper, whenever $L$ is known from the context, we use ${\bm f}|_{{\bmss X}={\bms d}}$ to denote ${\bm f}^{(L)}|_{{\bmss X}={\bms d}}$.
\begin{rmk}\label{rmk:nonzero}
Given a sparse sequence ${\bm f}^{(L)}|_{{\bmss X}={\bms d}}$, the number of non-zero elements is $2^{m-t}$. Hence, the sparsity of the sequence ${\bm f}^{(L)}|_{{\bmss X}={\bms d}}$ is given by
\begin{equation}
\mathcal{S}=\frac{L-2^{m-t}}{L}.
\end{equation}
\end{rmk}
\begin{eg}
Following the same notations given in \textit{Example~\ref{eg:gbf}}, let $t=1$, ${\bm X}=(x_2)$, and ${\bm d}=(0)$. The associated sequence ${\bm f}|_{{x_2}={0}}$ to the RGBF is ${\bm f}|_{{x_2}={0}}=(\xi^{0},\xi^{1},0,0,\xi^{0},\xi^{1},0,0)$. From (\ref{eq:front_end}), we have $k_0=0$ and $k_1=5$ and hence $L=6$. The truncated sequence is ${\bm f}^{(6)}|_{{x_2}={0}}=(\xi^{0},\xi^{1},0,0,\xi^{0},\xi^{1},\text{\st{0,0}})=$ $(\xi^{0},\xi^{1},0,0,\xi^{0},\xi^{1})$ with the sparsity $\mathcal{S}=\frac{6-4}{6}=\frac{1}{3}$.
\end{eg}
\section{Proposed Constructions of SCPs and Their Mates}\label{sec:SCP}
In this section, we will provide constructions of SCPs and their mates based on RGBFs.
\begin{thm}\label{thm:scp}
For integers $m$, $t$ with $t\leq m-1$, let $\pi$ be a permutation of the set $\{1,2,\ldots,m\}$ with $\pi(m)> \pi(\alpha)$ for $1\leq\alpha\leq t$. Let ${\bm X}=(x_{\pi(1)},x_{\pi(2)},\ldots,x_{\pi(t)})$, and binary sequence ${\bm d}=(d_1,d_2,\ldots,d_t)$.
Let us consider the RGBF given below:
\begin{equation}
\begin{aligned}
f|_{\bmss X={\bms d}}=\frac{q}{2}\Bigg(&\sum_{l=1}^{t-1}d_{l}d_{l+1}+\sum_{l=t+1}^{m-1}x_{\pi(l)}x_{\pi(l+1)}+d_{t}x_{\pi(t+1)}\Bigg)\\
&+\sum_{l=1}^{m}g_lx_{l}+g_0,
\end{aligned}
\end{equation}
where $g_l\in \mathbb{Z}_q$.
The pair
\begin{equation}\label{eq:th1}
({\bm C}_0,{\bm C}_1)=\left({\bm f}|_{\bmss X={\bms d}},({\bm f}+(q/2){\bm x}_{\pi(t+1)})|_{\bmss X={\bms d}}\right)
\end{equation}
is an $(L,Z,\mathcal{S})$-SCP with $L=\sum_{\alpha=t+1}^{m}2^{\pi(\alpha)-1}+1$, $Z=\sum_{\alpha=1}^{t}2^{\pi(\alpha)-1}+1$, and $\mathcal{S}=\frac{L-2^{m-t}}{L}$.\footnote{{The sparsity $\mathcal{S}$ is controlled by the sequence lengths and the number of restricted variables $t$. Since the proposed sequence length is associated with the restricted variables of RGBFs, for a fixed $t$, the sparsity can be controlled by selecting different restricted variables.}}
\end{thm}
\begin{IEEEproof}
Without loss of generality, we consider ${\bm d}={\bm 0}_{t}$ where ${\bm 0}_{t}$ is the all-zero vector of length $t$. Given the pair $({\bm C}_0,{\bm C}_1)$ generated from \textit{Theorem \ref{thm:scp}}, they can be written as ${\bm C}_k=(C_{k,0},C_{k,1},\ldots,C_{k,L-1})=\left({\bm f}+\frac{q}{2}\cdot k\cdot {\bm x}_{\pi(t+1)}\right)|_{{\bmss X}={\bm d}},~k=0,1$, where $C_{k,i}=0$ if $(i_{\pi(1)},i_{\pi(2)},\ldots,i_{\pi(t)})\neq (d_1,d_2,\ldots,d_{t})$ or $C_{k,i}=\xi^{c_{k,i}}$ if $i_{\pi(\alpha)}=d_{\alpha}$ for $\alpha=1,2,\ldots,t$. We need to show that C1 and C2 in (\ref{eg:scp}) hold.

In the first part, we start with checking C1, i.e.,
\begin{equation*}
\begin{aligned}
\rho({\bm C}_k,{\bm C}_{k'};u)&
=\sum_{i=0}^{L-1-u}C_{k,i+u}C^{*}_{k',i}\\&=
\begin{cases}
0, &0<u<Z,~ k=k'\\
0, &0\leq u<Z,~k\neq k'
\end{cases}
\end{aligned}
\end{equation*}
where $k,k'=0,1$. Let $j=i+u$ and also let $(j_1,j_2,\ldots,j_m)$ and $(i_1,i_2,\ldots,i_m)$ be binary representations of the integers $j=\sum_{l=1}^{m}j_{l}2^{l-1}$ and $i=\sum_{l=1}^{m}i_{l}2^{l-1}$, respectively. Note that ${C_{k,j}}C^{*}_{k',i}=0$ if $(j_{\pi(1)},j_{\pi(2)},\ldots,j_{\pi(t)})$ or $(i_{\pi(1)},i_{\pi(2)},\ldots,i_{\pi(t)})$ is not equal to $(d_1,d_2,\ldots,d_{t})$. Hence, throughout the proof, we only need to focus on the multiplications of non-zero elements, i.e., ${C_{k,j}}C^{*}_{k',i}\neq 0$ implying $(j_{\pi(1)},j_{\pi(2)},\ldots,j_{\pi(t)})=$ $(i_{\pi(1)},i_{\pi(2)},\ldots,i_{\pi(t)})=(d_1,d_2,\ldots,d_{t})$. Then three cases are considered below.

{\it Case 1 (i): $k=k'$.} In this case, we consider the aperiodic autocorrelation of ${\bm C}_k$. For $i_{\pi(\alpha)}=j_{\pi(\alpha)}$, $1\leq \alpha\leq t$, we show next that $i_{\pi(m)}=j_{\pi(m)}$ holds. Let us assume that $i_{\pi(m)}\neq j_{\pi(m)}$. Denote the set $W=\{w_0,w_1,w_2,\ldots,w_{\gamma}\}$ such that $\pi(m)=\pi(w_0)<\pi(w_{1})<\pi(w_{2})<\cdots<\pi(w_{\gamma})$ where $0\leq\gamma\leq m-t-1$. {Assume that $\kappa$ is the largest integer satisfying $i_{\pi(w_{\kappa})}\neq j_{\pi(w_{\kappa})}$
and $A = \{1, 2, \dots, m\} \setminus\{ 1, \dots, t, w_{\kappa+1}, \dots, w_{\gamma}\}$. That is to say,
$A$ contains all the indices $\alpha$ for which $i_{\pi(\alpha)} \neq j_{\pi(\alpha)}$.
% Then $i_{\pi(w)}= j_{\pi(w)}$ for $w\in W_1$.
Since $u=j-i>0$, we have $j_{\pi(w_{\kappa})}=1$ and $i_{\pi(w_{\kappa})}=0$. Then, we obtain
\begin{equation}
\begin{aligned}
u&=j-i=2^{\pi(w_{\kappa})-1}+\sum_{ \alpha \in A \setminus \{w_{\kappa}\}}(j_{\pi(\alpha)}-i_{\pi(\alpha)})2^{\pi(\alpha)-1}\\
&\geq 2^{\pi(w_{\kappa})-1}-\sum_{ \alpha \in A \setminus \{w_{\kappa}\}}2^{\pi(\alpha)-1}=\sum_{\alpha=1}^{t}2^{\pi(\alpha)-1}+1,
\end{aligned}
\end{equation}
since $\sum_{\alpha=1}^{t}2^{\pi(\alpha)-1}+\sum_{ \alpha \in A \setminus \{w_{\kappa}\}}2^{\pi(\alpha)-1}=2^{\pi(w_{\kappa})-1}-1$.} Obviously, this contradicts the assumption that $u<Z=\sum_{\alpha=1}^{t}2^{\pi(\alpha)-1}+1$. Thus, we have $i_{\pi(m)}=j_{\pi(m)}$. For simplicity, we denote $\sigma(\alpha)=\pi(m+1-\alpha)$. Then we have $i_{\sigma(1)}=j_{\sigma(1)}$. Assume that $\beta$ is the smallest integers such that $i_{\sigma(\beta)}\neq j_{\sigma(\beta)}$. Let $i'$ and $j'$ be integers different from $i$ and $j$, respectively, in the position $\sigma(\beta-1)$, i.e., $i'_{\sigma(\beta-1)}=1-i_{\sigma(\beta-1)}$ and $j'_{\sigma(\beta-1)}=1-j_{\sigma(\beta-1)}$. We can obtain
\begin{equation*}
\begin{aligned}
i',j'&\leq \sum_{\alpha=t+1}^{m}2^{\pi(\alpha)-1}-2^{\sigma(\beta-1)-1}+1<L.
%L-2^{\sigma(\beta-1)-1}
\end{aligned}
\end{equation*}
Then we have
\begin{equation*}
\begin{aligned}
&c_{k,i'}-c_{k,i}\\
&=\frac{q}{2}(i_{\sigma(\beta-2)}i'_{\sigma(\beta-1)}-i_{\sigma(\beta-2)}i_{\sigma(\beta-1)}\\
&+i'_{\sigma(\beta-1)}i_{\sigma(\beta)}-i_{\sigma(\beta-1)}i_{\sigma(\beta)})\\
&+g_{\sigma(\beta-1)}i'_{\sigma(\beta-1)}-g_{\sigma(\beta-1)}i_{\sigma(\beta-1)}\\
&\equiv \frac{q}{2}(i_{\sigma(\beta-2)}+i_{\sigma(\beta)})+g_{\sigma(\beta-1)}(1-2i_{\sigma(\beta-1)}) \pmod q.
\end{aligned}
\end{equation*}
Since $i_{\sigma(\beta-1)}=j_{\sigma(\beta-1)}$ and $i_{\sigma(\beta-2)}=j_{\sigma(\beta-2)}$, we possess
\begin{equation}
c_{k,j}-c_{k,i}-c_{k,j'}+c_{k,i'}\equiv \frac{q}{2}(i_{\sigma(\beta)}-j_{\sigma(\beta)})\equiv\frac{q}{2},
\end{equation}
implying $\xi^{c_{k,j}-c_{k,i}}+\xi^{c_{k,j'}-c_{k,i'}}=0.$

{\it Case 1 (ii): $k\neq k'$ and $u\neq 0$.} In this case, we show that $\rho({\bm C}_0,{\bm C}_1;u)=0$ for $0<u<Z_{A}$. Based on \textit{Case 1 (i)}, we know $i_{\pi(m)}=j_{\pi(m)}$ and let $\pi(\alpha)=\sigma(m+1-\alpha)$. Assume that $\beta$ is the smallest integer satisfying $i_{\sigma(\beta)}\neq j_{\sigma(\beta)}$. Let $i'$ and $j'$ be integers distinct from $i$ and $j$, respectively, in the position $\sigma(\beta-1)$. Following the similar arguments as given above, we can obtain $\xi^{c_{0,j}-c_{1,i}}+\xi^{c_{0,j'}-c_{1,i'}}=0$.

{\it Case 1 (iii): $k\neq k'$ and $u\neq 0$.} In this case, we prove $\rho({\bm C}_0,{\bm C}_1;0)=0$. From (\ref{eq:th1}), we have $\rho({\bm C}_0,{\bm C}_1;0)= \sum_{i=0}^{L-1}C_{0,i}C^{*}_{1,i}$ where $C_{0,i}C^{*}_{1,i}=\xi^{\frac{q}{2}i_{\pi(t+1)}}$ for $i_{\pi(\alpha)}=d_{\alpha}$, $1\leq\alpha\leq t$, and $i_{\pi(t+1)}$ is the $\pi(t+1)$-th bit of the binary representation of $i$.
According to \textit{Remark \ref{rmk:nonzero}}, each sequence has $2^{m-t}$ non-zero elements so that we can find $2^{m-t-1}$ pairs fulfilling $\xi^{c_{0,i}-c_{1,i}}=\xi^{q/2}=-1$ and another $2^{m-t-1}$ pairs such that $\xi^{c_{0,i}-c_{1,i}}=\xi^{0}=1$. Therefore, we have $\rho({\bm C}_0,{\bm C}_1;0)=0$.

From \textit{Case 1 (i)} to \textit{Case 1 (iii)}, we see C1 of (\ref{eq:SCP}) holds.

Now we consider condition C2. That is to say, we need to show that
\begin{equation*}
\rho({\bm C}_0;u)+\rho({\bm C}_1;u)=0
\end{equation*}
for $0<u<L$. Taking $i_{\pi(\alpha)}=j_{\pi(\alpha)}=d_{\alpha}$ for $\alpha=1,2,\ldots,t$, the following two cases are considered.

{\it Case 2 (i):} Suppose that $i_{\pi(t+1)}\neq j_{\pi(t+1)}$. We can obtain
\begin{equation*}
c_{0,j}-c_{0,i}-c_{1,j}+c_{1,i}=\frac{q}{2}(i_{\pi(t+1)}-j_{\pi(t+1)})\equiv \frac{q}{2} \pmod q,
\end{equation*}
meaning that $\xi^{c_{0,j}-c_{0,i}}+\xi^{c_{1,j}-c_{1,i}}=0.$

{\it Case 2 (ii):} Considering $i_{\pi(t+1)}= j_{\pi(t+1)}$, we assume $\beta$ is the smallest integer which fulfills $i_{\pi(\beta)}\neq j_{\pi(\beta)}$. Likewise, let $i'$ and $j'$ be integers different from $i$, $j$, respectively, in the position $\pi(\beta-1)$. Following the similar arguments as given in \textit{Case 1 (i)}, we obtain
\begin{equation*}
\xi^{c_{0,j}-c_{0,i}}+\xi^{c_{0,j'}-c_{0,i'}}+\xi^{c_{1,j}-c_{1,i}}+\xi^{c_{1,j'}-c_{1,i'}}=0,
\end{equation*}
which completes the proof.
\end{IEEEproof}
\begin{rmk}
If we consider $t=0$ in \textit{Theorem \ref{thm:scp}}, the pair $({\bm C}_0,{\bm C}_1)$ is reduced to a GCP of length $2^m$ which is the so-called Golay-Davis-Jedweb (GDJ) pair \cite{Golay_RM}.
\end{rmk}
\begin{rmk}
According to \textit{Theorem \ref{thm:scp}}, the sequence length
and the ZCZ width satisfy
$$
L + Z = \sum_{\alpha=t+1}^{m}2^{\pi(\alpha)-1}+1 +
\sum_{\alpha=1}^{t}2^{\pi(\alpha)-1}+1 = 2^m + 1.
$$
Since $\{\pi(1),\pi(2),\ldots,\pi(m)\}=\{1,2,\ldots,m\}$ with $\pi(m)>\pi(\alpha)$ for $1\leq\alpha\leq t$, by taking different values of $t$ and choices of $\pi$, we can flexibly generate SCPs with arbitrary length in the range $[2^{m-1}+1, 2^m]$. In addition, we point out that there is a one-to-one correspondence between the sequence length $L$ and $t$, permutation $\pi.$

In Table \ref{table}, we list some examples of the proposed SCPs of lengths from 15 to 35 except for the lengths
16, 20, 26, 32 for binary GCPs, where the corresponding ZCZ widths and sparsities are given as well. It can be observed that by restricting different variables in \textit{Theorem \ref{thm:scp}}, flexible lengths of SCPs are obtained.

\end{rmk}
% \begin{IEEEproof}

% \end{IEEEproof}

\begin{table*}[t!]
\caption{SCPs of Lengths up to 35\label{table}}
\begin{center}
\scriptsize
\setlength\extrarowheight{4pt}
\begin{tabular}
{|c||c|c|c|c|c|c|c|c|c|c|c|c|c|c|c|c|c|c|c|c|c|c|c|c|}\hline
Length              & 15   & 17   &18   &19   &21   &22   &23    &24    &25   &27   &28 & 29   & 30   &31   &33   &34   &35                 \\ \hline
$m$                 &4     &5 &5 &5 &5 &5 &5 &5 &5 &5 &5  &5     &5 &5 &6 &6 &6                            \\ \hline
\begin{tabular}[c]{@{}l@{}}Restricting\\  variables \end{tabular}        &$x_1$     & \begin{tabular}[c]{@{}l@{}}$x_1,x_2$\\  $x_3,x_4$ \end{tabular} &\begin{tabular}[c]{@{}l@{}}$x_2,x_3$\\  $x_4$ \end{tabular} &\begin{tabular}[c]{@{}l@{}}$x_1,x_3$\\  $x_4$ \end{tabular} &\begin{tabular}[c]{@{}l@{}}$x_1,x_2$\\  $x_4$ \end{tabular} &\begin{tabular}[c]{@{}l@{}}$x_2$\\  $x_4$ \end{tabular} &\begin{tabular}[c]{@{}l@{}}$x_1$\\  $x_4$ \end{tabular} &$x_4$ &\begin{tabular}[c]{@{}l@{}}$x_1,x_2$\\  $x_3$ \end{tabular} &\begin{tabular}[c]{@{}l@{}}$x_1$\\  $x_3$ \end{tabular} &$x_3$ &\begin{tabular}[c]{@{}l@{}}$x_1$\\  $x_2$ \end{tabular}     & $x_2$ &$x_1$ &\begin{tabular}[c]{@{}l@{}}$x_1,x_2$\\  $x_3,x_4$\\$x_5$ \end{tabular} &\begin{tabular}[c]{@{}l@{}}$x_2,x_3$\\  $x_4,x_5$ \end{tabular} &\begin{tabular}[c]{@{}l@{}}$x_1,x_3$\\  $x_4,x_5$ \end{tabular}                             \\ \hline
ZCZ width                &2     &16 &15 &14 &12 &11 &10 &9 &8 &6 &5 &4     &3 &2 &32 &31 &30                            \\ \hline
%$ZCZ_{\text{ratio}}$                 &$\frac{2}{15}$    &$\frac{16}{17}$ &$\frac{15}{18}$ &$\frac{14}{19}$ &$\frac{12}{21}$ &$\frac{11}{22}$ &$\frac{10}{23}$ &$\frac{9}{24}$ &$\frac{8}{25}$ &$\frac{6}{27}$ &$\frac{5}{28}$ &$\frac{4}{29}$    &$\frac{3}{30}$ &$\frac{2}{31}$ &$\frac{32}{33}$ &$\frac{31}{34}$ &$\frac{30}{35}$                            \\[0.15cm] \hline
 Sparsity           &$\frac{7}{15}$ &$\frac{15}{17}$ &$\frac{14}{18}$ &$\frac{15}{19}$ &$\frac{17}{21}$ &$\frac{14}{22}$ &$\frac{15}{23}$ &$\frac{8}{24}$ &$\frac{21}{25}$ &$\frac{19}{27}$ &$\frac{12}{28}$ &$\frac{21}{29}$ &$\frac{24}{30}$ &$\frac{15}{31}$ &$\frac{31}{33}$ &$\frac{30}{34}$ &$\frac{31}{35}$
\\[0.15cm] \hline
\end{tabular}
\end{center}
 \end{table*}
\begin{eg}\label{eg:scp}
Taking $q=4$, $m=5$, $t=2$, and $\pi=(1,3,2,4,5)$, let ${\bm X}=(x_1,x_3)$ and ${\bm d}=(d_{1},d_2)=(0,0)$. The RGBF is
\begin{equation*}
f|_{{\bmss X}={\bms d}}=2\cdot(x_2x_4+x_4x_5+d_1d_2+d_2x_2)+3x_2.
\end{equation*}
According to \textit{Theorem \ref{thm:scp}}, the pair
\begin{equation*}
({\bm C}_0,{\bm C}_1)=({\bm f}|_{{\bmss X}={\bms d}},({\bm f}+2{\bm x}_2)|_{{\bmss X}={\bms d}})
\end{equation*}
is a $(27,6,19/27)$-SCP where
\begin{equation*}
    {{\bm C}_0=(\xi^{0}0\xi^{3}00000\xi^{0}0\xi^{1}00000\xi^{0}0\xi^{3}00000\xi^{2}0\xi^{3})}
\end{equation*}
and
\begin{equation*}
 {{\bm C}_1=(\xi^{0}0\xi^{1}00000\xi^{0}0\xi^{3}00000\xi^{0}0\xi^{1}00000\xi^{2}0\xi^{1})}.
\end{equation*}
In Fig. \ref{fig:scp_sum}, we can see ${\bm C}_0$ and ${\bm C}_1$ have zero AACS for any non-zero time-shift. Besides, in Fig. \ref{fig:scp_aper}, the ZCZ width of 6 of the aperiodic autocorrelations of ${\bm C}_0$ and cross-correlations of ${\bm C}_0$ and ${\bm C}_1$ are illustrated.
\begin{figure}[ht!]
	\centering
	\includegraphics[width = 3.5in]{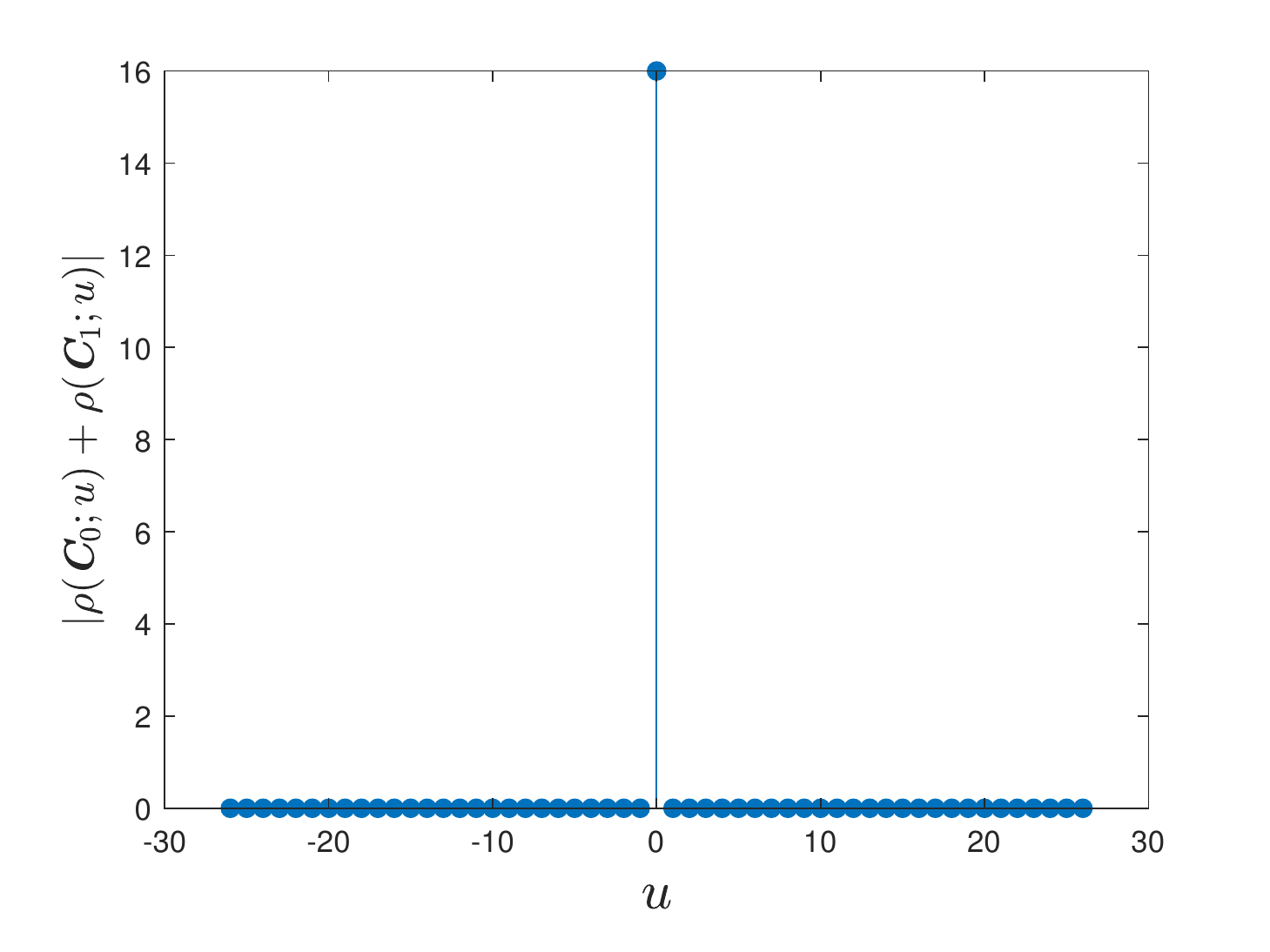}
	\caption{Sum of aperiodic autocorrelations of ${\bm C}_0$ and ${\bm C}_1$ in \textit{Example \ref{eg:scp}}.}
    \label{fig:scp_sum}	
\end{figure}
\begin{figure}[ht!]
	\centering
	\includegraphics[width = 3.5in]{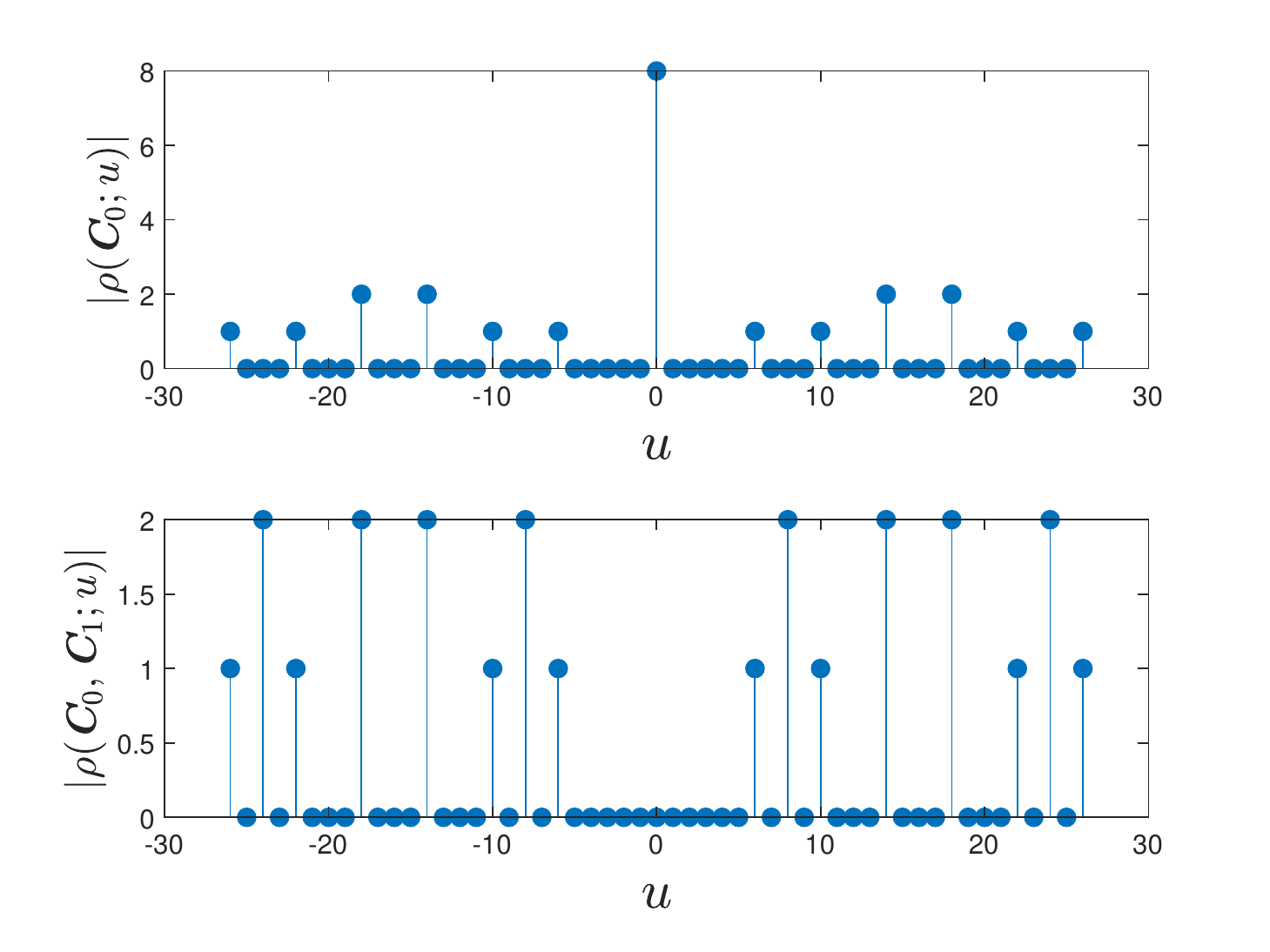}
	\caption{Aperiodic autocorrelations of ${\bm C}_0$ and cross-correlations of ${\bm C}_0$ and ${\bm C}_1$ in \textit{Example \ref{eg:scp}}.}
    \label{fig:scp_aper}	
\end{figure}
\end{eg}

Next, we present a mate construction of an SCP based on $\textit{Theorem~\ref{thm:scp}}$.
\begin{thm}\label{thm:scm}
Let $\pi(m-1)>\pi(\alpha)$ for $1\leq\alpha\leq t$, and $0\leq t\leq m-2$ in \textit{Theorem \ref{thm:scp}}, and $({\bm C_0},{\bm C}_1)$ be the SCP given in (\ref{eq:th1}). Then the pair $({\bm S_0},{\bm S}_1)$ given by
\begin{equation*}
\begin{aligned}
% &({\bm S_0},{\bm S}_1)\\
&\left(\left({\bm f}+\frac{q}{2}{\bm x}_{\pi(m)}\right)\Big|_{{\bmss X}={\bms d}},\left({\bm f}+\frac{q}{2}{\bm x}_{\pi(t+1)}+\frac{q}{2}{\bm x}_{\pi(m)}\right)\Big|_{{\bmss X}={\bms d}}\right)
\end{aligned}
\end{equation*}
is a sparse mate of $({\bm C_0},{\bm C}_1)$.

\end{thm}
\begin{IEEEproof}
For the two $(L,Z,\mathcal{S})$-SCPs $({\bm C}_0,{\bm C}_1)$ and $({\bm S}_0,{\bm S}_1)$, we show that (\ref{eq:scm}) holds. Likewise, we only consider the condition $i_{\pi(\alpha)}=j_{\pi(\alpha)}=d_{\alpha}$, where $1\leq \alpha\leq t$. In the first part, we prove
\begin{equation*}
\begin{aligned}
&\rho({\bm C}_k,{\bm S}_{k'};u)=0,~k,k'=0,1
%&\rho({\bm C}_0,{\bm S}_0;u)+\rho({\bm C}_1,{\bm S}_1;u)=0,~\text{for}~0\leq u<L.
\end{aligned}
\end{equation*}
for $0\leq u<Z$. Following similar notations as given in the proof of \textit{Theorem \ref{thm:scp}}, let ${\bm S}_{k'}=(S_{k',0},S_{k',1},\ldots,S_{k',L-1})$ where $S_{k',i}=\xi^{s_{k',i}}$ if $(i_{\pi(1)},i_{\pi(2)},\ldots,i_{\pi(t)})= \bm d$ and $S_{k',i}=0$ otherwise. We consider two cases below.

{\it Case 1 (i):} Given $u>0$, $0\leq t\leq m-2$, we already have $i_{\pi(m)}=j_{\pi(m)}$. In this case, we next show that $i_{\pi(m-1)}=j_{\pi(m-1)}$. To proceed, let us assume that $i_{\pi(m-1)}\neq j_{\pi(m-1)}$. Denote the set $\hat{W}=\{\hat{w}_0,\hat{w}_1,\ldots,\hat{w}_{\gamma}\}$ such that $\pi(m-1)=\pi(\hat{w}_0)<\pi(\hat{w}_{1})<\pi(\hat{w}_{2})<\cdots<\pi(\hat{w}_{\gamma})$ where $0\leq\gamma\leq m-t-1$. Assume that $\lambda$ is the largest integer satisfying $i_{\pi(\hat{w}_{\lambda})}\neq j_{\pi(\hat{w}_{\lambda})}$ and $\hat{A} = \{1, 2, \dots, m\} \setminus\{ 1, \dots, t, \hat{w}_{\lambda+1}, \dots, \hat{w}_{\gamma}\}$.
{ If $\pi(\hat{w}_{\lambda})<{\pi(m)}$, we have
\begin{equation*}
\begin{aligned}
u=j-i&=2^{\pi(\hat{w}_{\lambda})-1}+\sum_{\alpha \in \hat{A} \setminus \{\hat{w}_{\lambda}\}}^{m}(j_{\pi(\alpha)}-i_{\pi(\alpha)})2^{\pi(\alpha)-1}\\
&\geq 2^{\pi(\hat{w}_{\lambda})-1}-\sum_{\alpha \in \hat{A} \setminus \{\hat{w}_{\lambda}\}}^{m}2^{\pi(\alpha)-1}=\sum_{\alpha=1}^{t}2^{\pi(\alpha)-1}+1,
\end{aligned}
\end{equation*}
which violates the assumption $u<\sum_{\alpha=1}^{t}2^{\pi(\alpha)-1}+1$. Similar results can be obtained if $\pi(\hat{w}_{\lambda})>{\pi(m)}$. Hence, we have $i_{\pi(m-1)}=j_{\pi(m-1)}$. }
Similarly we denote  $\pi(\alpha)=\sigma(m+1-\alpha)$ and assume $\beta$ is the smallest integers such that $i_{\sigma(\beta)}\neq j_{\sigma(\beta)}$ where $\beta> 2$. Then let $i'$ and $j'$ be integers different from $i$ and $j$, respectively, in the position $\sigma(\beta-1)$. Similar results can be obtained as provided in \textit{Case 1 (i)} of the proof of \textit{Theorem \ref{thm:scp}}. Hence, we possess
$\xi^{c_{k,j}-s_{k',i}}+\xi^{c_{k,j'}-s_{k',i'}}=0.$

{\it Case 1 (ii):} In this case, we show
\begin{equation*}
\rho({\bm C}_k,{\bm S}_{k'};0)=\sum_{i=0}^{L-1}C_{k,i}S_{k',i}=0,
%=\sum_{i=0}^{L-1}\xi^{\frac{q}{2}((k-k')i_{\pi(t+1)}-i_{\pi(m)})}
\end{equation*}
where $C_{k,i}S_{k',i}= \xi^{c_{k,i}-s_{k',i}}$ and
\begin{equation}\label{eq:mate_0}
c_{k,i}-s_{k',i}=\frac{q}{2}\left((k-k')i_{\pi(t+1)}-i_{\pi(m)}\right)
\end{equation}
for $i_{\pi(\alpha)}=d_{\alpha}$, $1\leq \alpha\leq t$.
Clearly, (\ref{eq:mate_0}) can be regarded as the linear combination of the terms $i_{\pi(t+1)}$ and $i_{\pi(m)}$. Hence, $\sum_{i=0}^{L-1}C_{k,i}S_{k',i}=0.$

From \textit{Case 1 (i)} and \textit{Case 1 (ii)}, we can see that the cross-correlation of sequences ${\bm C}_k$ and ${\bm S}_{k'}$ is zero within the ZCZ $Z$. Then in the second part, we need to show
\begin{equation*}
\rho({\bm C}_0,{\bm S}_{0};u)+\rho({\bm C}_1,{\bm S}_{1};u)=0.
\end{equation*}
Three cases are considered below.

{\it Case 2 (i):}  For $u>0$, if $\pi(t+1)\neq \pi(t+1)$, we follow a similar derivation in \textit{Case 2 (i)} of the proof of \textit{Theorem \ref{thm:scp}} and have $\xi^{c_{0,j}-s_{0,i}}+\xi^{c_{1,j}-s_{1,i}}=0$.

{\it Case 2 (ii):}  For $u>0$, assume $i_{\pi(t+1)}= j_{\pi(t+1)}$. Let $\beta$ be the smallest integers such that $i_{\pi(\beta)}\neq  j_{\pi(\beta)}$. Similar results can be obtained as given in \textit{Case 2 (ii)} of the proof of \textit{Theorem 1}. We have $\xi^{c_{0,j}-s_{0,i}}+\xi^{c_{0,j'}-s_{0,i'}}+\xi^{c_{1,j}-s_{1,i}}+\xi^{c_{1,j'}-s_{1,i'}}=0$.

{\it Case 2 (iii):} Finally, it only suffices to validate that $\rho({\bm C}_0,{\bm S}_{0};0)+\rho({\bm C}_1,{\bm S}_{1};0)=0$. By recalling the \textit{Case 1 (ii)} of this proof, we complete the proof.
\end{IEEEproof}
\begin{eg}\label{eg:scm}
Let us follow the same notations given in \textit{Example \ref{eg:scp}}. According to \textit{Theorem \ref{thm:scm}}, we let $\pi(4)=4$ and obtain the pair $({\bm S_0},{\bm S}_1)=(({\bm f}+2{\bm x}_5)|_{{\bmss X}={\bms d}},({\bm f}+2{\bm x}_2+2{\bm x}_5)|_{{\bmss X}={\bms d}})$ given by
\begin{equation*}
{{\bm S}_0=(\xi^{0}0\xi^{3}00000\xi^{0}0\xi^{1}00000\xi^{2}0\xi^{1}00000\xi^{0}0\xi^{1})}
\end{equation*}
and
\begin{equation*}
 {{\bm S}_1=(\xi^{0}0\xi^{1}00000\xi^{0}0\xi^{3}00000\xi^{2}0\xi^{3}00000\xi^{0}0\xi^{3})}.
\end{equation*}
In Fig. \ref{fig:scm_aper}, we illustrate the aperiodic cross-correlations of ${\bm C}_0$ and ${\bm S}_0$ and cross-correlations of ${\bm C}_0$ and ${\bm S}_1$. The ZCZ width is 6. Besides, it can be validated that they have zero cross-correlation sums for every time-shift.
\begin{figure}[ht!]
	\centering
	\includegraphics[width = 3.5in]{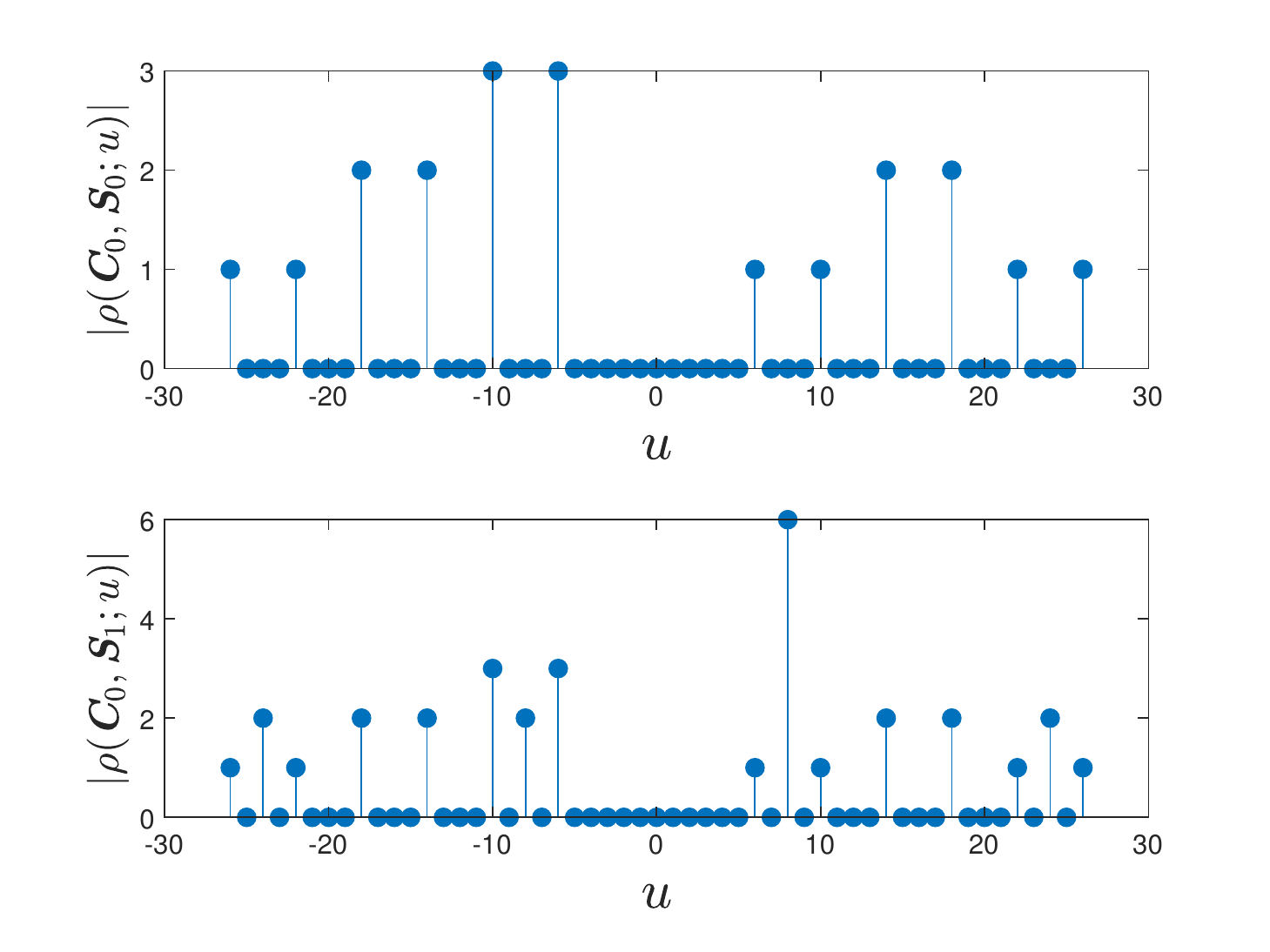}
	\caption{Aperiodic cross-correlations of ${\bm C}_0$ and ${\bm S}_0$ and cross-correlations of ${\bm C}_0$ and ${\bm S}_1$ in \textit{Example \ref{eg:scm}}.}
    \label{fig:scm_aper}	
\end{figure}
\end{eg}
\section{Conclusion}\label{sec:conclusion}
In this paper, we have introduced a novel class of sequence pairs called SCPs in which every SCP has zero AACS at every non-zero time-shift and the aperiodic autocorrelations and cross-correlations of the two constituent sequences are zero within a zone. Thanks to the use of RGBFs, we have introduced a direct construction of SCPs with arbitrary lengths and controllable sparsity levels. We have further constructed the mutually orthogonal mate of an SCP by considering cross-correlation properties between the two distinct SCPs.

%A family of  2-D $((2^m,2^nL),(2^m,Z))$-ZCAPs can be obtained by concatenations, recursively, based on 1-D $(L,Z)$-ZCPs. The size of 2-D ZCZ depends on the width of ZCZ of the 1-D kernel ZCPs. In addition, Theorem \ref{thm:flex} provides a construction of $((2L_1,L_2),(L_1,Z))$-ZCAPs with flexible sizes where $L_1$ and $L_2$ are not necessarily powers of 2. Moreover, we modify the construction of 1-D ZCPs based on Kronecker product in \cite{ZCP-1st} and then, extend it to construct a 2-D $((L_1L_3,L_2L_4),(L_1Z_3,L_2Z_4))$-ZCAPs based on an $L_1\times L_2$ GCAP and an $((L_3,L_4),(Z_3,Z_4))$-ZCAP. Although only binary arrays are taken into considerations in this paper, the proposed constructions can be modified for nonbinary cases in the future.
%\begin{appendices}
%%--------------------------------APPENDIX A--------------------------------%
%
%\section{Proof of Theorem \ref{thm:mate}}\label{apxB}
%%
%%\section*{Acknowledgment}
%%
%%We are indebted to Michael Shell for maintaining and improving
%%\texttt{IEEEtran.cls}.
%\end{appendices}
\bibliographystyle{IEEEtran}
\bibliography{IEEEabrv,2D_ZCP}
%%%%%%
%% To balance the columns at the last page of the paper use this
%% command:
%%
%\enlargethispage{-1.2cm}
%%
%% If the balancing should occur in the middle of the references, use
%% the following trigger:
%%
\IEEEtriggeratref{3}
%%
%% which triggers a \newpage (i.e., new column) just before the given
%% reference number. Note that you need to adapt this if you modify
%% the paper.  The "triggered" command can be changed if desired:
%%
%\IEEEtriggercmd{\enlargethispage{-20cm}}
%%
%%%%%%

\end{document}